\documentstyle[12pt]{article}
\def\le{\langle}
\def\re{\rangle}

\def\K{{\cal K}}

\def\1{\mbox{I\hspace{-.15em}1}}

\def\R{{\rm I\hspace{-.15em}R}}

\def\C{\hspace{3pt}{\rm l\hspace{-.47em}C}}

\def\b{\begin{equation}}
\def\e{\end{equation}}
\def\bee{\begin{enumerate}}
\def\eee{\end{enumerate}}
\def\re{\rangle}

\def\b{\begin{equation}}
\def\e{\end{equation}}
\def\bd{\begin{displaystyle}}
\def\ed{\end{displaystyle}}
\def\ba{\begin{array}}
\def\ea{\end{array}}

\def\bee{\begin{enumerate}}
\def\eee{\end{enumerate}}

\def\bes{\begin{eqnarray*}}
\def\ees{\end{eqnarray*}}
\def\be{\begin{eqnarray}}
\def\ee{\end{eqnarray}}

\def\le{\langle}
\def\re{\rangle}

\def\R{{\rm I\hspace{-.15em}R}}

\def\C{\hspace{3pt}{\rm l\hspace{-.47em}C}}

\title{Quantum Linear Gravity in de Sitter Universe \\ On Gupta-Bleuler vacuum state}

\author{M. Enayati, S. Rouhani, M.V. Takook\thanks{e-mail:
takook@razi.ac.ir}}

\date{\today}

\begin{document}

\maketitle 
\centerline{$^1$Department of Physics, Razi
University, Kermanshah, IRAN}

\begin{abstract}

Application of Krein space quantization to the linear gravity in de Sitter space-time have constructed on Gupta-Bleuler vacuum state, resulting in removal of infrared divergence and preserving de Sitter covariant. By pursuing this path, the non uniqueness of vacuum expectation value of the product of field operators in curved space-time disappears as well. Then the vacuum expectation value of the product of field operators can be defined properly and uniquely.

\end{abstract}

\vspace{0.5cm}
{\it Proposed PACS numbers}: 04.62.+v, 03.70+k, 11.10.Cd, 98.80.H
\vspace{0.5cm}

\section{Introduction} 

One of the challenging goals in theoretical physics is constructing a proper covariant quantization of the
gravitational field. The gravitational red-shift leads us to the
conclusion that gravity could be explained at least partially through geometry  \cite{mtw73}. Two dominant views in
geometry have been utilized for this purpose. In the first
perspective, geometry is completely defined by the Riemannian
curvature tensor $R_{\mu\nu\rho\sigma}$ or equivalently by the
metric tensor $g_{\mu\nu}$ (due to the metric compatibility: $
\bigtriangledown_\mu g_{\rho\nu}= 0$, and torsion free condition
$\Gamma_{\mu\nu}^{\rho}=\Gamma_{\nu \mu}^{\rho}$). The gravitational field can be explained by
the Einstein field equation, confirming relatively well by the
experimental tests in the solar system scale. Moreover, the
gravitational field is described by an irreducible rank-2 symmetric
tensor field $g_{\mu\nu}$.

The other perspective for geometry discards the metric compatibility
and torsion free condition \cite{hehl,vsl10}. In this schema, geometry is
defined by the connection coefficients and metric tensors
$(\Gamma_{\mu\nu}^{\rho}, g_{\mu\nu})$-or equivalently by physical
quantities of Riemann curvature tensor, torsion tensor and metric
tensor ($R_{\mu\nu\rho\sigma}, T_{\;\;\mu\nu}^{\rho}, g_{\mu\nu}$). In
this view the gravitational field can not be described by merely an
irreducible rank-2 symmetric tensor field $g_{\mu\nu}$. Instead the
physical triplet $R_{\mu\nu\rho\sigma}, T_{\;\;\mu\nu}^{\rho}$ and
$g_{\mu\nu}$ ($RTg$) determines the behaviour of the gravitational
field. A special and simple case of this approach is Weyl geometry \cite{fulton,di73}, in which the linear approximation of the gravitational field is described by a rank-3 mix-symmetry tensor field \cite{bfh83,tatafa10,tapeta12}.

Because of infra-red and ultraviolet divergences, the quantum field theory (QFT) is problematic. These divergences not only may violate the
principle of covariance and the gauge symmetry, but prevent
the calculation of the expectation value of physical quantities.
Exertion of certain additional methods such as ''re-normalization''
have been able us to resolve just one part of the above anomalies. These
methods, however, have twofold problems: (1) they are not a part of
the fundamental theory of quantum mechanics and (2) they are not
able to solve many problems of quantum field theories such as quantum
gravity. Distortion of  concepts such as time and causality are
the first obstacles in the process of quantization of gravitational
field. Both of them have been removed in the
''background field'' method, but quantum linear gravity is not
re-normalizable in this method.

It should be noted that the problem of divergences are not inherent
in gravitational field theory, but a defect inherited from quantum
field theory. In other words, even in the Minkowsky space-time, the
quantization of the field theory results in divergences which have to be
removed arbitrarily, if possible, in order to preserve the
compatibility with actual physical measurement. In the case of the
general relativity, however, one can not eliminate the divergences,
i.e. the theory is non-renormalizable. We have discussed that this
anomaly in the QFT disappears in Krein space quantization in the one-loop approximation.

The recent cosmological observations are strongly in favour of positive acceleration for present universe and thus most suitably represented by de Sitter space-time. In other words, de Sitter space-time is an excellent choice for representing the background space-time of our universe. 
The linear quantum gravity in de Sitter space was
studied thoroughly by Iliopoulos et al. \cite{il,anilto}. They have shown that the pathological large-distance behavior (infra-red divergence) of the
graviton propagator on a de Sitter background does not manifest
itself in the  quadratic part of the effective action  in the
one-loop approximation \cite{il,anilto}. This means that the pathological behaviour of the graviton propagator may be gauge dependent and so should not
appear in an effective way as a physical quantity.
The linear gravity (the traceless rank-2 ``massless'' tensor field) on de
Sitter ambient space formalism was built rudimentary from the minimally coupled
scalar field \cite{ta97}. It has been shown that the application of Krein space
quantization to the minimally coupled scalar field in de Sitter 
space has resulted in removal of infrared and ultraviolet divergences 
and henceforth naturally maintained the principle of causality  \cite{ta97,dere,gareta1}.
Construction of linear gravity in de Sitter universe through Krein space 
quantization which will present in this paper, pursuing a similar path for 
the earlier works \cite{gareta1,ta3}. A notable consequence of this construction 
is absence of divergence in Green function at large distances, resulting in 
removal of infra-red divergence \cite{gareta1,ta3}. This method was applied in various area of QFT and/or QED 
where all resulted in natural renormalization of the 
solutions \cite{fotaza12,ta02,ta05,knrt01,rt11}.

Although negative norm states appear in our method, by imposing the
following conditions they are effectively removed and the unitarity of theory is preserved:
\begin{itemize}
\item[i)] The first condition is the "reality condition" in which
the negative norm states do not appear in the external legs of the
Feynmann diagram. This condition guarantees that the negative norm
states only appear in the internal legs and in the disconnected
parts of the diagram. \item[ii)]The second condition is that the S
matrix elements must be renormalized in the following form:
$$S_{if}\equiv\mbox{probability amplitude}=\frac{<\mbox{physical states}
, in|\mbox{physical states}, out>}{<0,in|0,out>}.$$
This condition eliminates the negative norm states in the
disconnected parts.
\end{itemize}

In previous methods the choice of vacuum state directly affected
the expectation values of energy momentum tensor. In the present
method, however, vacuum expectation values are independent of the choice
of modes. Although the expectation value of the
energy momentum tensor for physical states are dependent on the
choice of modes, the expectation value of vacuum states remains
uniquely the same.

\setcounter{equation}{0}
\section{Krein space quantization }

Let us briefly describe our quantization of the minimally coupled
massless scalar field in de Sitter space, which can be identified by a 4-dimensional hyperboloid embedded in 5-dimensional Minkowskian space-time:
     \b X_H=\{x \in \R^5 ;x^2=\eta_{\alpha\beta} x^\alpha
 x^\beta =-H^{-2}\},\;\; \alpha,\beta=0,1,2,3,4, \e
where $\eta_{\alpha\beta}=$diag$(1,-1,-1,-1,-1)$ and $H$ is Hubble parameter. For simplicity from now on we take $H=1$ in some equations and insert it again, whenever it is needed. The de Sitter
metrics is \b  ds^2=\eta_{\alpha\beta}dx^{\alpha}dx^{\beta}|_{x^2=-H^{-2}}=
g_{\mu\nu}^{dS}dX^{\mu}dX^{\nu},\;\; \mu=0,1,2,3,\e where $X^\mu$
are 4 space-time intrinsic coordinates. Any geometrical
object in this space can be written either in terms of the four local coordinates $X^\mu$ or the five global coordinates $x^\alpha$with the condition $(2.1)$ (ambient space formalism). The choices of bounded global
coordinates $(X^\mu,\;\mu=0,1,2,3)$ is well suited for
compactified de Sitter space, namely S$^3 \times{\rm S}^1$ by the metric
\b \label{dscor}  ds^2=g^{dS}_{\mu\nu}dX^\mu dX^\nu=\frac{1}{H^2 \cos^2\rho}
(d\rho^2-d\alpha^2-\sin^2\alpha\,
d\theta^2-\sin^2\alpha\sin^2\theta \,d\phi^2).\e

The minimally coupled massless scalar field is defined by  \b  \Box_H
\varphi(x)=0,\e where $\Box_H$ is the Laplace-Beltrami operator on de
Sitter space. In ambient space formalism the solution can be written in terms of dS plane wave \cite{brmo} $$\phi(x)=(Hx\cdot \xi)^\sigma ,$$ where $\xi
$ lies on the positive null cone $ {\cal C}^+ = \{ \xi \in \R^5;\;\; \xi\cdot\xi=0, \xi^0>0
\}$ and $\sigma=0$ or $-3$ is the homogeneous degree of the scalar field. Due to the zero mode problem (or constant solution $\sigma=0)$, one cannot
construct a covariant quantum field in the usual manner \cite{al}. An approach to the solution of this problem has been achieved by introducing a specific Krein QFT \cite{gareta1}. Here
we review the construction of minimally coupled massless field
in de Sitter space, which will be used in linear quantum gravity in section $4$. We return to
intrinsic coordinates in order to restore the covariance. The solution to
the field equation $(2.4)$
 reads in the coordinate system (\ref{dscor}) (for $L\neq 0$ )
\cite{gareta1}: \b\phi_{Llm}(x)=X_{L}(\rho)Y_{Llm}(\Omega)\equiv \phi_{k},\e with
\b X_{L}(\rho)=\frac{H}{2}[2(L+2)(L+1)L]^{-\frac{1}{2}} \left(L
e^{-i(L+2)\rho}+(L+2)e^{-iL\rho}\right).\e And for $L= 0$, we have
$$\phi_{000}=\phi_g+\frac{1}{2}\phi_s,\;\phi_g=\frac{H}{2\pi}\;\;
\phi_s=-i\frac{H}{2\pi}[\rho+\frac{1}{2}\sin 2\rho ].$$
The $y_{Llm}(\Omega)$'s are the hyper-spherical harmonics. As proved by Allen
\cite{al}, the covariant canonical quantization
 procedure with positive norm states fails in this case. The Hilbert space generated by the positive modes, including the zero mode ($\phi_{000}$), is
 not de Sitter invariant \cite{al},
 $${\cal H}=\{\sum_{k \geq 0}\alpha_k\phi_k;\;
 \sum_{k \geq 0}|\alpha_k|^2<\infty\}.$$
  This means that it is not closed under the action of the de~Sitter group generators.
 Nevertheless, one can obtain a fully covariant quantum field by adopting
 a new construction \cite{dere,gareta1}. In order to obtain a fully covariant
 quantum field, we have added all the conjugate modes to the previous ones.
 Consequently, we have to deal with an orthogonal sum of a positive and
 negative inner product space, which is closed under an indecomposable
 representation of the de~Sitter group. Now, the decomposition of the field operator into positive and negative norm parts is:
   \b \varphi(x)=\frac{1}{\sqrt{2}}\left[ \varphi_p(x)+\varphi_n(x)\right],\e
 where
 \b \varphi_p(x)=\sum_{k\geq 0} a_{k}\phi_{k}(x)+H.C.,\;\;
  \varphi_n(x)=\sum_{k \geq 0} b_{k}\phi^*_k(x)+H.C..\e
 The positive mode $\varphi_p(x)$ is the scalar field as was used by Allen.
 The crucial departure from the standard QFT based on CCR lies in the
 following form:
 \b   [a_{k},a^{\dag} _{k'}]= \delta_{kk'},\;\;
  \;\;[b_{k},b^{\dag} _{k'}]= -\delta_{kk'}.\e
The ground state is defined as the Gupta-Bleuler vacuum state:
$$ a_{k}|GB>=0,\;\; \;\;
   b_{k}|GB>=0.$$
A direct consequence of this construction is the positivity of the energy  {\it i.e.}
$$\le\vec k|T_{00}|\vec k\re\geq0,$$ for any physical state $|\vec
k\re$ (those built from repeated action of the $a^{\dag} _{k}$'s on
the vacuum state),
$$|\vec k\re=|k_1^{n_1}\ldots k_j^{n_j}\re=\frac{1}{\sqrt{n_1!\ldots n_j!}}
\left(a_{k_1}^{\dag}\right)^{n_1}
\ldots\left(a_{k_j}^{\dag}\right)^{n_j}|GB\re.$$
 This quantity vanishes if and only if $|\vec
k\re=|GB\re$. Therefore the ``normal ordering'' procedure for
eliminating the ultraviolet divergence in the vacuum energy, which
appears in the usual QFT is not needed \cite{gareta1}. Another
consequence of this quantization is a covariant two-point function, which
is free of any infrared divergence \cite{ta3}. 

\setcounter{equation}{0}
\section{Scalar field in general curved space}

In general curved space-time the scalar field equation is
\begin{equation}
\Box \varphi +m^2\varphi +\zeta R\varphi =0. \label{eq:KG}
\end{equation}
Here $R$ is the Riemann scalar curvature, and $\zeta$ is a coupling
constant. The inner product of a pair of their solutions is defined by \cite{bida}
\begin{equation}
(\phi_1,\phi_2)=i\int (\phi^*_2 \,
{\mathop{\partial_\mu}\limits^\leftrightarrow } \,\phi_1) d\Sigma^\mu,
\end{equation}
where $d\Sigma^\mu = d\Sigma\, n^\mu$. $d\Sigma$ is the
volume element in a given space-like hyper-surface, and $n^\mu$ is
the time-like unit vector normal to this hyper-surface.  Let $\{ \phi_k \}$ be a set of solutions of positive norm states of Eq.
(\ref{eq:KG}),
\b
(\phi_k,\phi_{k'})=\delta _{kk'}, \;\;
(\phi_k^*,\phi_{k'}^*)= -\delta _{kk'}, \;\;
(\phi_k,\phi_{k'}^*)=  0, \label{eq:ortho}
\e
then $\{ \phi^*_k \}$ will be a set of solutions of negative
norm states.  We have proved that the set $\{ \phi_k \}$ is not a complete set of solutions for minimally coupled scalar field in de Sitter space
\cite{gareta1}. Thus, in general, $\{ \phi_k, \phi^*_k \}$ form a complete set of
solutions of the wave equation, in terms of which we may expand as an
arbitrary solution. The field operator $\varphi$ in Krein space
quantization can be written as a sum of positive and negative field
operators:
$$ \varphi = \frac{1}{\sqrt{2}}\left(\varphi_p+\varphi_n\right)= \frac{1}{\sqrt{2}}\left[\sum_k a_k \phi_k +
a^\dagger_k \phi^*_k + \sum_k b_k \phi^*_k +
b^\dagger_k \phi_k\right],$$
or
$$
\varphi = \sum_k \left[\frac{a_k+b^\dagger_k}{\sqrt{2}} \phi_k+
\frac{a^\dagger_k+b_k}{\sqrt{2}} \phi^*_k\right],
$$
where $$[a_k, a^\dagger_{k'}] = \delta_{kk'},\;\; [b_k,
b^\dagger_{k'}] = -\delta_{kk'},$$ and the other commutation
relations are zero.  The Gupta-Bleuler vacuum state 
$|GB^{\phi}\rangle$ is defined such as \b a_k |GB^{\phi}\rangle=0,\;\;\;\;\; b_k |GB^{\phi}\rangle=0,\e  and the physical and un-physical states are respectively:
$$a_k^\dag |BG^{\phi}\rangle=|k^{\phi}\rangle,\;\;\;\;\; b_k^\dag |GB^{\phi}\rangle= |\bar k^{\phi}\rangle.$$ 

In curved space-time, there is,
in general, no unique choice of the mode solution $\{ \phi_k , \phi^*_k\}$, and hence there exists no
unique notion of the vacuum state. This means that the
notion of ``particle'' becomes ambiguous and as a consequences cannot be properly defined in general curved space time. Let $\{ F_j , F^*_j\}$ be another set of
solutions of the field equation.  We may choose these sets of solutions to be orthonormal
\begin{equation}
(F_{j},F_{j'})=\delta _{jj'},\;\;
(F_{j}^*,F_{j'}^*)= -\delta _{jj'},\;\;
 (F_{j},F_{j'}^*)= 0.  \label{eq:ortho2}
\end{equation}
We may expand the $\phi$-modes in terms of the F-modes:
\begin{equation}
\phi_k=\sum\limits_j {(\alpha _{kj}}F_j + \beta _{kj}F_j^*).
\end{equation}
Inserting this expansion into the orthogonality relations, Eqs.
(\ref{eq:ortho}) and (\ref{eq:ortho2}), leads to the conditions
\begin{equation}
\sum\limits_j {(\alpha _{kj}\alpha _{k'j}^*-\beta
_{kj}\beta_{k'j}^*) = \delta _{kk'}}, \label{eq:alphabeta}
\end{equation}
and
\begin{equation}
 \sum\limits_j (\alpha _{kj}\alpha _{k'j}-\beta _{kj}\beta _{k'j})=0.
\end{equation}
The inverse expansion is
\begin{equation}
F_j=\sum\limits_k {(\alpha _{kj}^*}\phi_k - \beta _{kj}\phi_k^*).
\end{equation}

The field operator $\varphi$ in Krein space quantization may be expanded in
terms of either of the two sets: $\{ \phi_k, \phi^*_k\}$ or $\{ F_j , F^*_j\}$:
\begin{equation}
  \varphi =\sum_k\left[\frac{a_k+b^\dagger_k}{\sqrt{2}} \phi_k +
\frac{a^\dagger_k+b_k}{\sqrt{2}} \phi^*_k\right]
  =\sum_j \left[\frac{c_j+d^\dagger_j}{\sqrt{2}} F_j +
\frac{c^\dagger_j+d_j}{\sqrt{2}} F^*_j\right].
\end{equation}
The $a_k$ and $a_k^\dagger$ are annihilation and creation operators of physical state,
respectively, in the $\phi$-modes, whereas the $c_j$ and $c_j^\dagger$
are the corresponding operators for the F-modes. $b_k$,
$b_k^\dagger$ and  $d_j$, $d_j^\dagger$ are annihilation and
creation operators of the un-physical states in their respective
mode solutions. The $\phi$-vacuum state, which is defined by $(3.4)$ describes the situation when no particle (and un-physical state) is
present in this state. The F-vacuum state is defined by
$c_j|GB^F\rangle=0, \;\; d_j|GB^F\rangle=0 \; \forall j,$ and describes the situation where
no particle (and un-physical state) is present. Noting that $a_k = (\varphi,\phi_k)$ and $c_j
= (\varphi,F_j)$, we may expand the two sets of creation and
annihilation operator in terms of one another as
\begin{equation}
a_k=\sum\limits_j (\alpha _{kj}^*c_j-\beta _{kj}^* c_j^\dagger),\;\;
b_k^\dagger=\sum\limits_j (\alpha _{kj}^*d_j^\dagger-\beta _{kj}^*
d_j), \label{eq:Bogo1}
\end{equation}
and
\begin{equation}
c_j=\sum\limits_k (\alpha _{kj} a_k + \beta _{kj}^* a
_k^\dagger),\;\; d_j^\dagger=\sum\limits_k (\alpha _{kj} b_k^\dagger
+ \beta _{kj}^* b _k). \label{eq:Bogo2}
\end{equation}
This is a Bogoliubov transformation, and the $\alpha_{kj}$ and
$\beta_{kj}$ are called the Bogoliubov coefficients.

In Krein space quantization, it is possible to describe the physical phenomenon of particle creation by a time-dependent gravitational field similar to the usual quantization. The physical number
operator $N_j^c = c_j^\dagger
c_j$ counts particles in the F-modes. If any $\beta_{kj}$ coefficients are non-zero,
\begin{equation}
\langle GB^{\phi}|N_j^c |GB^{\phi}\rangle = \langle GB^{\phi}|c_j^\dagger c_j |GB^{\phi}\rangle
       = \sum\limits_k |{\beta _{kj}}|^2,
\end{equation}
i.e. if any
mixing of positive and negative frequency solutions are presented, then
particles are created by the gravitational field.

One of the most fundamental problems of QFT in curved space-time is that
the vacuum expectation value of physical quantities (such as
$T_{\mu\nu}$) is depend on the choice of vacuum states. This is
direct consequence of non-uniqueness of the vacuum states. In the
Krein space quantization in-spite of non-uniqueness of the vacuum
states, which results in ambiguity of the concept of particle, the choice of vacuum states does not affect the vacuum
expectation value of physical quantities. These quantities are defined uniquely by:
\b \langle GB^{\phi}|T_{\mu\nu} |GB^{\phi}\rangle=0=\langle GB^F|T_{\mu\nu}
|GB^F\rangle. \e
However, the expectation value of $T_{\mu\nu}$ on physical states depends on the choice of mode 
\b \langle k^{\phi}|T_{\mu\nu} |k^{\phi}\rangle\neq\langle j^F|T_{\mu\nu}
|j^F\rangle. \e It is noted again that Krein space quantization was proved to remove the divergences of QFT in the one-loop approximation\cite{gareta1,fotaza12,knrt01,rt11} and as well the linear quantum gravity, which will be considered in the next section.

\setcounter{equation}{0}
\section{Linear quantum gravity }

The massless spin-2 field in dS space-time, $h_{\mu\nu}(X)$, can be considered as: \b
g_{\mu\nu}=g^{dS}_{\mu\nu}+h_{\mu\nu},\e where $g^{dS}_{\mu\nu}$ is
the gravitational de Sitter background and $h_{\mu\nu}$ is the gravitational waves or the fluctuation
part. For ``massless'' tensor fields $h_{\mu\nu}(X)$, the wave equation which
propagate on de Sitter space can be derived through a variation of the action integral \b  S=-\frac{1}{16 \pi G}\int
(R-2 \Lambda)\sqrt{-g}\, d^4X,\e where $G$ is the Newtonian constant
and  $\Lambda$ is the cosmological constant. Application of the
variational calculus leads to the field equation \b
R_{\mu\nu}-\frac{1}{2}Rg_{\mu\nu}+\Lambda g_{\mu\nu}=0. \e 
The wave equation, which obtain in the linear approximation, is
\cite{fr}: $$  -(\Box_H+2H^2)h_{\mu\nu}-(\Box_H+H^2)
g_{\mu\nu}h'-2 \nabla_{(\mu}\nabla^{\rho}h_{\nu)\rho}$$ \b
+g_{\mu\nu}\nabla^{\lambda}\nabla^{\rho}h_{\lambda\rho}
+\nabla_{\mu}\nabla^{\nu}h'=0,\e where $h'=h_{\mu}^{\mu}$.
$\nabla^\nu$  is the covariant derivative on dS space. As usual, two
indices inside parentheses mean that they are symmetrized, {\it
i.e.} $T_{(\mu\nu)}=\frac{1}{2}(T_{\mu\nu}+T_{\nu\mu})$. The field
equation $(4.4)$ is invariant under the following gauge transformation
\b h_{\mu\nu} \longrightarrow
h_{\mu\nu}^{gt}=h_{\mu\nu}+2\nabla_{(\mu}A_{\nu)},\e 
where $A_{\nu}$ is an arbitrary vector field. One can choose a general
family of gauge conditions \b  \nabla^{\mu}h_{\mu \nu}=l
\nabla_{\nu}h',\e where $l$ is an arbitrary constant. If the value of
$l$ is set to be $\frac{1}{2}$, the relation between unitary
representation and the field equation not only becomes clearly
apparent but also reduces to a very simple form. The tensor field notation $\K_{\alpha\beta}(x)$ (ambient space formalism) is adapted to establish this relation. 

The tensor field $h_{\mu\nu}(X)$ is locally
determined by the tensor field $\K_{\alpha\beta}(x)$: \b  h_{\mu\nu}(X)=\frac{\partial x^{\alpha}}{\partial
X^{\mu}}\frac{\partial x^{\beta}}{\partial
X^{\nu}}\K_{\alpha\beta}(x(X)),\;\; \K_{\alpha\beta}(x)
 =\frac{\partial X^{\nu}}{\partial x^{\alpha}}
\frac{\partial X^{\mu}}{\partial x^{\beta}}h_{\mu\nu}(X(x)). \e The
field $\K_{\alpha\beta}(x)$ which is defined on de Sitter space-time is a homogeneous function in the $\R^5$-variables
$x^{\alpha}$: \b
x^{\alpha}\frac{\partial }{\partial
x^{\alpha}}\K_{\beta\gamma}(x)=x\cdot\partial \K_{\beta\gamma}
(x)=\sigma \K_{\beta\gamma}(x), \e where $\sigma$ is an arbitrary degree of homogeneity. It also satisfies the conditions
of transversality \cite{di35} \b x\cdot\K(x)=0,\mbox{ \it i.e.
}x^\alpha \K_{\alpha\beta}(x)=0,\mbox{ and } x^\beta
\K_{\alpha\beta}(x)=0 . \e

In order to obtain the wave equation for the tensor field $\K$, we
must use the tangential (or transverse) derivative $\bar \partial$
on de Sitter space \b \bar \partial_\alpha=\theta_{\alpha
\beta}\partial^\beta=
\partial_\alpha  +H^2x_\alpha x\cdot\partial,\;\;\;x\cdot\bar \partial=0,\e
where $\theta_{\alpha \beta}=\eta_{\alpha \beta}+H^2x_{\alpha}x_{
\beta}$ is the transverse projector. To express tensor field in ambient space formalism, transverse projection is defined \cite{ga, tata10} $$
(Trpr \K)_{\alpha_1 \cdots \alpha_l}\equiv
\theta_{\alpha_1}^{\beta_1}
\cdots\theta_{\alpha_l}^{\beta_l}\K_{\beta_1 \cdots \beta_l}\;.$$
The transverse projection guarantees the transversality in each
index. Therefore, the covariant derivative of a tensor field,
$T_{\alpha_1....\alpha_n}$, in the ambient space formalism becomes
$$ Trpr \bar\partial_\beta \K_{\alpha_1 ..... \alpha_n}\equiv
\nabla_\beta T_{\alpha_1....\alpha_n}\equiv \bar
\partial_\beta
T_{\alpha_1....\alpha_n}-H^2\sum_{i=1}^{n}x_{\alpha_i}T_{\alpha_1..\alpha_{i-1}\beta\alpha_{i+1}..\alpha_n},$$
so we have  \b
\nabla_{\mu}h_{\nu\rho}\longrightarrow
\theta_{\alpha}^{\alpha'}\theta_{\beta}^{\beta'}
\theta_{\gamma}^{\gamma'}
\partial_{\alpha'}\K_{\beta'\gamma' }.\e

The field equation for $\K$ from $(4.4)$ is shown to be \cite{fr,tata10,gata1}
\b B[(Q_2+6)\K(x)+D_2\partial_2 \cdot\K]=0,  \e where operator $B$ is defined as 
$BT=T-\frac{1}{2}\theta T'$ with $T':=\eta^{\alpha \beta} T_{\alpha
\beta}$. $Q_2$ is the Casimir operator of the de Sitter group and the subscript $2$ in $Q_2$ shows that the carrier space encompasses second rank tensors \cite{gata1}. The operator $D_2$ is the generalized gradient \b
D_2K=H^{-2}{\cal S}(\bar
\partial-H^2x)K,\e
where ${\cal S}$ is the symmetrizer operator. The generalized divergence is defined by ($\partial_2\cdot $): \b \partial_2\cdot
\K=\partial^T\cdot\K-\frac{1}{2}H^2D_1\K'=\partial \cdot\K- H^2 x
\K'-\frac{1}{2} \bar  \partial \K',\e where
$\partial^T\cdot\K=\partial\cdot\K-H^2x\K'$ is the transverse
divergence and $D_1=H^{-2}\bar \partial$. One can invert the operator $B$ and hence write the equation $(4.12)$ in the form \b  (Q_2+6)\K(x)+D_2\partial_2 \cdot\K=0.\e
This equation is gauge invariant, {\it i.e. }
$\K^{gt}=\K+D_2\Lambda_g$ is also a solution of $(4.15)$ for any vector field
$\Lambda_g$ satisfying the conditions: $\partial \cdot \Lambda_g=0=x\cdot \Lambda_g$. The equation $(4.15)$ can be derived from
the Lagrangian density \b {\cal
L}=\frac{H^2}{2}\K..(Q_2+6)\K+(\partial_2 \cdot\K)^2,\e
where $..$ is a shortened notation for total contraction. The gauge fixing condition $(4.6)$ reads in our notations as
\b  \partial_2\cdot\K=(l-\frac{1}{2})\bar \partial \K'.\e For the value of $l=1/2$, chosen by Christensen and Duff \cite{chdu}, we have \b  \partial_2\cdot\K=0.\e

Similar to the flat space QED, gauge fixing is accomplished by adding to $(4.16)$ a gauge fixing term: \b {\cal
L}=\frac{H^2}{2}\K..(Q_2+6)\K+(\partial_2
\cdot\K)^2+\frac{1}{\alpha}  (\partial_2 \cdot\K)^2.\e The variation
of ${\cal L} $ then leads to the equation \cite{gaha} \b
(Q_2+6)\K(x)+cD_2\partial_2 \cdot\K=0.\e where
$c=\frac{1+\alpha}{\alpha}$ is a gauge fixing term. Actually, the
simplest choice of $c$ is not zero, as it will be shown later.

In the general gauge condition $(4.17)$ the gauge
fixing Lagrangian is \b {\cal
L}=\frac{H^2}{2}\K..(Q_2+6)\K+\frac{1}{2}(\partial_2
\cdot\K)^2+\frac{1}
{\alpha}\left[\partial_2\cdot\K-(l-\frac{1}{2})\bar
\partial \K'\right]^2.\e
The field equation which derives from this Lagrangian becomes
$$(Q_2+6)\K(x)+D_2\partial_2 \cdot\K
+\frac{1}{\alpha}\left[D_2\partial_2 \cdot\K+(l-\frac{1}{2})^2\eta
(\bar \partial)^2\K'-(l-\frac{1}{2})(D_2\bar \partial \K'-{\cal
S}\bar \partial\partial_2 \cdot\K)\right] =0. $$ Clearly this equation is
more complicated than (4.20) obtained by the choice of $l=\frac
{1}{2}$. In the following we shall work with the choice
$l=\frac{1}{2}$ only.

\subsection{Gupta-Bleuler triplet}

The appearance of the Gupta-Bleuler triplet seems to be a universal phenomenon in gauge theories,
and indeed a crucial element for field quantization \cite{bifrhe}. The ambient space formalism
allows us to exhibit this triplet for the linear gravity in exactly the
same manner as the electromagnetic field quantization. Let us now define the Gupta-Bleuler triplet $V_g \subset V \subset V_{c}$, carrying the indecomposable representation of de Sitter group:

\begin{itemize}

\item[-] The space  $V_{c}$ is the space  of all square integrable solutions of the field
equation (4.20), including negative norm states. It is
c dependent so that one can actually adopt an optimal value of c
which eliminates logarithmic divergent \cite{gaha}. In
the next section, we will show  that this particular value is
$c=\frac{2}{5}$. More generally, for a spin s field,
$c=2/(2s+1)$ is the proper value  \cite{ga}.

\item[-] It contains a subspace $V$ of solutions,
satisfying the divergencelessness condition $\partial\cdot{\cal K}=0$. This subspace
$V$ is not invariant under the action of the de Sitter group generators. In view of Eq.
(4.20), it  is obviously $c$ independent as well.

\item[-] The subspace $V_g$ of $V$ consists of the pure gauge solutions of the form
${\cal K}_g=D_2K_g$ with $\partial\cdot{\cal K}_g=0$. These are orthogonal to every element in
$ V$, including themselves. This subspace
$V_g$ is also not invariant under the action of the de Sitter group generators.
\end{itemize}

The pure gauge vector field $K_g$, which correspond to the vector space $V_{g}$, satisfy the field equation \cite{ta14}: 
$$ \left(Q_1+6\right)K_g(x)=0,\;\;\; x\cdot K_g=0=\partial \cdot K_g,$$
$Q_1$ is the Casimir operator of the de Sitter group and the subscript $1$ reminds us that the carrier spaces do encompass vector field. The vector field $K_g$ can be transformed by the representation $U^{1,2}(g)$, which does not belong to any UIR of de Sitter group. The vector field $K(x)=\partial_2\cdot{\cal K}$ belonging to the space
$V_{c}/V$, satisfy the following field equation \cite{ta14}:
$$ \left(Q_1+6\right)K(x)=0,\;\;\; x\cdot K=0=\partial \cdot K.$$ 
Similar to the above pure gauge field, $K$ can be transformed by the representation $U^{1,2}(g)$. The tensor field ${\cal K}$, which correspond to the subspace $V$, satisfies the field equation:
$$ (Q_2+6)\K(x)=0,\;\; \partial\cdot{\cal K}=0.$$
The physical states belong to the quotient space $V/V_{g}$. The physical states can be transformed by the spin-$2$ unitary irreducible representations of the dS group  ($\Pi^+_{2,2} \bigoplus \Pi^-_{2,2}$) that admit a Minkowskian massless
spin-$2$ interpretation \cite{ta14}.

\subsection{dS-field solution}

A general solution of equation $(4.20)$ can be constructed by a
scalar field and two vector fields. Let us introduce a tensor field
$\K$ in terms of a five-dimensional constant vector $Z=(Z_\alpha)$
and a scalar field $\phi_1$ and two vector fields $K$ and $K_g$ by
putting \b \K=\theta\phi_1+ {\cal S}\bar Z_1K+D_2K_g,\e where $\bar
Z_{1\alpha}=\theta_{\alpha\beta} Z^{\;\beta}_1$. Substituting  $\K$
into $(4.20)$, using the commutation rules and following algebraic
identities \cite{tata10,gaha,berotata06,derotata08}: \b Q_2\theta \phi=\theta Q_0\phi,\e \b
\partial_2 \cdot \theta\phi=-H^2D_1\phi,  \e
\b Q_2D_2K_g=D_2Q_1K_g,\e \b \partial_2 \cdot D_2K_g = -(Q_1+6)K_g,\e \b
Q_2 {\cal S}\bar Z K={\cal S}\bar Z (Q_1-4)K-2H^2D_2x\cdot
ZK+4\theta Z\cdot K
,\e \b \partial_2 \cdot {\cal S}\bar Z_1K=\bar Z_1 \partial_1 \cdot
K-H^2D_1Z\cdot K-H^2xZ\cdot K+Z\cdot(\bar
\partial+5H^2x)K, \e
we find that
$K$ obeys the wave equation $$( Q_1+2)K+cD_1\partial\cdot  K=0,\;
x\cdot K=0.$$ $Q_0$ is the Casimir operators of the de Sitter group which act on scalar field. If we impose the
supplementary condition $\partial \cdot K=0$, we get \b ( Q_1+2)K=0,\;
x \cdot K=0=\partial\cdot K.\e

The vector field $K$ as a consequences of their conditions could be
transformed as a representation of de Sitter group  \cite{gata00, gagarota08}.
The further following choice of condition \b \phi_1
=-\frac{2}{3}Z_1\cdot K, \e results in $$
(Q_1+6)K_g =\frac{c}{2(c-1)}H^2 D _1\phi_1 + \frac{2-5c}{1-c} H^2 x\cdot
Z_1 K+$$ \b \frac{c}{1-c} (H^2 x Z_1\cdot K - Z_1\cdot \bar\partial K),\e where $K_g$ can be also determined in terms of $K$. Then the scalar field $\phi_1$ satisfies the following wave equation
 \b Q_0 \phi_1=0 ,\e
where $\phi_1$ is a ``massless'' minimally coupled scalar field. If we chose $c=\frac{2}{5}$, then we get the simplest form for $K_g$.
\b  K_g =\frac{1}{9}\left(H^2 x Z_1\cdot K- Z_1\cdot\bar\partial
K+\frac{2}{3}H^2 D_1 Z_1\cdot K\right).\e In conclusion, if we know the
vector field $K$, we will also know the tensor field $\K$.

The vector field $K$ in de Sitter space was explicitly studied in previous papers \cite{gata00, gagarota08}. It can be written in terms of two scalar field $\phi_2$ and $\phi_3$: \b K_{\alpha}=\bar Z_{2\alpha}\phi_2+D_{1\alpha}\phi_3.\e
Applying $K_\alpha$ to equation $(4.29)$ result in the following equations
$$ Q_0\phi_2=0,$$ \b \phi_3 =-\frac{1}{2} [Z_2\cdot\bar \partial
\phi_2+2H^2 x\cdot Z_2\phi_2],\e where $\phi_2$ is also a ``massless''
minimally coupled scalar field. Therefore, we can construct the
tensor field $\K$ in terms of two ``massless'' minimally coupled
scalar fields $\phi_1$ and $\phi_2$. But both fields are related by
$(4.30)$. Therefore, one can write \b \K_{\alpha \beta}(x)={\cal
D}_{\alpha \beta}(x,\partial,Z_1,Z_2)\phi,\;\; \phi=\phi_2,\e where
$$ {\cal D}(x,\partial,Z_1,Z_2)=\left(-\frac{2}{3}\theta Z_1\cdot+{\cal
S}\bar Z_1+\frac{1}{9 }D_2 (H^2 xZ_1\cdot-Z_1\cdot\bar \partial
+\frac{2}{3}H^2 D_1 Z_1\cdot)\right)$$ \b\left( \bar Z_{2}-\frac{1}{2}
D_{1}(Z_2\cdot\bar
\partial+2H^2x\cdot Z_2)\right),\e
and $\phi$ is a ``massless'' minimally coupled scalar field, which was given by equation $(2.5)$. The
solution $(4.36)$ can be written as \b \K_{\alpha \beta}(x)={\cal
D}_{\alpha \beta}(x,\partial,Z_1,Z_2)\phi_{Llm}(x)={\cal D}_{\alpha
\beta}(x,\partial,Z_1,Z_2)X_L(\rho)y_{Llm}(\Omega).\e 
$Z_1$ and $Z_2$ are two constant vectors. We choose them in such a way that in
the limit $H=0$, one can obtain the polarization tensor in the
Minkowskian space $\epsilon_{\mu\nu}(k)$ \cite{we}: \b \lim_{H \rightarrow 0}{\cal D}_{\alpha
\beta}(x,\partial,Z_1,Z_2)\frac{X_L(\rho)y_{Llm}(\Omega)}{H\sqrt H}
\equiv \epsilon_{\mu\nu}(k)\frac{e^{ ik.X}}{\sqrt{k_0}} ,\e where
\b k^{\mu}\epsilon_{\mu\nu}(k)-\frac{1}{2}k_{\nu}\epsilon_{\nu}^{\nu}(k)=0,\;\;\; \epsilon_{\mu\nu}(k)=\epsilon_{\nu\mu}(k),\;\;k^{\nu}k_{\nu}=0.\e

Finally, we can write the solution under the form \b \K_{\alpha
\beta}(x)={\cal D}_{\alpha\beta}^{\lambda}(x,\partial)
\phi_{Llm}(\rho,\Omega)\equiv {\cal
E}_{\alpha\beta}^{\lambda}(\rho,\Omega,Llm)\phi_{Llm}(\rho,\Omega),\e
where ${\cal E}$ is the generalized polarization tensor and the
index $\lambda$ runs on all possible polarization states. The
explicit form of the polarization tensor is actually not important
here \cite{ta14}. Indeed, one can find the two-point function by just using the
recurrence formula $(4.22)$. In order to determine the generalized
polarization tensor ${\cal E}$, we let the projection operator ${\cal
D}$ acts on the scalar field $(2.5)$ and takes the Minkowskian limit $(4.39)$.

The solution ($4.22$) is traceless $\K'=0$. Let us now consider the
pure trace part (conformal sector) \b \K^{pt}=\frac{1}{4}\theta
\psi. \e Implementing this to equation $(4.20)$, we obtain $$
(Q_0+6)\psi+\frac{c}{2}Q_0\psi=0,$$ or \b
(Q_0+\frac{12}{2+c})\psi=0.\e On the other hand, any scalar field
corresponding to the discrete series representation of the dS group
obeys the equation \b  (Q_0+n(n+3))\psi=0.\e Hence, we see that the
value  $c=\frac{2}{5} $ does not correspond to any unitary irreducible
representations of the dS group. For $c=\frac{2}{5}$ the conformal sector satisfies:
$$ (\Box_H-5H^2)\psi=0.$$ Difficulties arise when we attempt to
quantize these fields where the mass square has negative values (conformal
sector with $c> -2$ and discrete series with $n>0$). The two-point
functions for these fields have a pathological large-distance
behavior. If we choose $c < -2$, this  behavior for the
conformal sector will disappear (although it is still present in the trace-less
part). In the following sections, the advantage of Krein space
quantization vividly shows itself where the pathological large-distance
behavior disappears in the trace-less
part. In the next section, we shall merely consider the traceless part, since it bears the physical states.

\subsection{Two-point function}

The quantum field theory of the ``massive'' spin-$2$
field (divergenceless and traceless) have been already constructed from the
Wightman two-point function ${\cal W}^\nu_{\alpha\beta\alpha'\beta'}$  \cite{gata1}: \b {\cal W}^\nu_{\alpha\beta
\alpha'\beta'}(x,x')=\langle
\Omega,\K_{\alpha\beta}(x)\K_{\alpha'\beta'}(x')\Omega  \rangle ,
\;\;\alpha,\beta, \alpha',\beta'=0,1,..,4,\e where $x,x'\in X_H$ and
$\mid \Omega  \rangle$ is the Fock vacuum state which is equivalent to the Bunch-Davies vacuum state. We have found that
this function can be written under the form \b {\cal
W}^\nu_{\alpha\beta \alpha'\beta'}(x,x')=D_{\alpha\beta
\alpha'\beta'}(x,x';\nu){\cal W}^\nu(x,x'), \e where $D_{\alpha\beta \alpha'\beta'}(x,x';\nu)$ is the
projection tensor and ${\cal
W}^\nu(x,x')$ is the Wightman two-point function for the massive
scalar field. Of course, we could crudely replace $\nu$
(principal-series parameter) by $\pm \frac{3}{2}i$ in order to get the ``massless'' tensor field which associated
to linear quantum gravity in dS space. However, this procedure leads to appearance of
two types of manageable singularities in the definition of the
two-point function ${\cal
W}^{\pm \frac{3}{2}i}_{\alpha\beta \alpha'\beta'}(x,x')$. The first one appears in the projection tensor
$D_{\alpha\beta \alpha'\beta'}(x,x';\pm \frac{3}{2}i)$ which could be removed by fixing
the gauge $(c=\frac{2}{5})$. The other appears in the scalar two-point function ${\cal W}^{\pm \frac{3}{2}i}(x,x')$, corresponding to
the minimally coupled scalar field, that is removed by Krein space quantization.

Let us briefly recall the required conditions for the ``massless''
bi-tensor two-point function ${\cal W}$, which is defined by
     \begin{equation}  {\cal W}_{\alpha\beta\alpha'\beta'}(x,x')=\langle
     GB|\K_{\alpha\beta}(x)\K_{\alpha'\beta'}(
     x')|GB\rangle,     \end{equation}
where $|GB \rangle$ is Gupta-Bleuler vacuum state \cite{gareta1}. These functions entirely  encode
 the theory of the generalized free fields on dS
space-time $X_H$. They have to satisfy the following requirements:
\begin{enumerate}
\item[a)] {\bf Indefinite sesquilinear form} for any test function $f_{\alpha \beta} \in {\cal D}(X_H)$, we
have an indefinite sesquilinear form that is defined by
\begin{equation} \int _{X_H \times X_H} f^{*\alpha\beta}(x)
{\cal W}_{\alpha\beta\alpha'\beta'}(x,x')f^{\alpha'\beta'}
(x')d\sigma(x)d\sigma(x'),\end{equation} where $ f^*$ is the
complex conjugate of $f$ and $d\sigma (x)$ denotes the
dS-invariant measure on $X_H$. ${\cal D}(X_H)$ is the space of rank-2 tensor
functions $C^\infty$ with compact support in $X_H$.
\item[b)] {\bf Locality} for every space-like separated pair $(x,x')$, {\it i.e.} $x\cdot
x'>-H^{-2}$,
\begin{equation}
{\cal W}_{\alpha\beta \alpha'\beta'}(x,x')={\cal
W}_{\alpha'\beta'\alpha\beta }(x',x) .\end{equation}
 \item[c)] {\bf Covariance} for all $g\in SO_0(1,4)$,
     \begin{equation} 
(g^{-1})^{\gamma}_{\alpha}(g^{-1})^{\delta}_{\beta} {\cal
W}_{\gamma\delta \gamma'\delta'} (g x,g x')g^{\gamma'}_{\alpha'}
g^{\delta'}_{\beta'}= {\cal W}_{\alpha\beta \alpha'\beta'}(x,x')
      .\end{equation}
\item[d)] {\bf Index symmetrizer}
\b {\cal W}_{\alpha\beta \alpha'\beta'}(x,x')={\cal W}_{\beta
\alpha \beta'\alpha '}(x,x').\e
\item[e)] {\bf Transversality}
\begin{equation}
x^\alpha {\cal W}_{\alpha\beta
\alpha'\beta'}(x,x')=0=x'^{\alpha'}{\cal W}_{\alpha\beta
\alpha'\beta'}(x,x') .\end{equation}
\item[f)] {\bf Tracelessness}
\b {\cal W}^\alpha_{\;\;\;\alpha \alpha'\beta'} (x,x')=0={\cal
W}_{ \alpha \beta \alpha '}^{\;\;\;\;\;\;\;\;\alpha'}(x,x').\e
\end{enumerate}

The two-point function ${\cal W}_{\alpha\beta
\alpha'\beta'}(x,x')$, which is a solution of the wave equation
$(4.20)$ with respect to $x$ and $x'$, can be found simply in
terms of the scalar two-point function. Let us try the following formulation for a transverse two-point
function:
\b\label{4.2} {\cal W}_{\alpha\beta
\alpha'\beta'}(x,x')=\theta_{\alpha\beta}
\theta'_{\alpha'\beta'}{\cal W}_0(x,x')+{\cal S}{\cal
S}'\theta_{\alpha}\cdot \theta'_{\alpha'}{
W}_{1\beta\beta'}(x,x')+D_{2\alpha}D'_{2\alpha'}{
W}_{g\beta\beta'}(x,x'),\e note that $D_2D'_2=D'_2D_2$ and ${
W}_{1}$ and ${ W}_{g}$ are transverse bi-vector two-point
functions which will be identified later. The calculation of ${\cal W}_{\alpha\beta
\alpha'\beta'}(x,x')$ could be initiated from either $x$ or $x'$, without
any difference. This means, other choices result in the same
equation for ${\cal W}_{\alpha\beta \alpha'\beta'}(x,x')$. By imposing this function to obey equation $(4.20)$,
with respect to $x$, it is easy to show that: \b \left\{
\begin{array}{ll}
          (Q_0+6)\theta'{\cal W}_0=-4{\cal S}'\theta'\cdot {\cal W}_{1},\;\;\;\;\;\;\;\;\;\;\;\;\;\;\;\;\;\;\;
          \;\;\;\;\;\;\;\;\;\;\;\;\;\;\;\;\;\;\;\;$(I)$\\
          \\(Q_1+2){\cal W}_{1}=0, \;\;\;\;\;\;\;\;\;\;\;\;\;\;\;\;\;\;\;\;\;\;\;\;\;\;\;\;\;\;\;\;\;
          \;\;\;\;\;\;\;\;\;\;\;\;\;\;\;\;\;\;\;\;\;\;$(II)$\\
          \\(Q_1+6)D'_2{\cal W}_g=\frac{c}{1-c}H^2D_1\theta'{\cal W}_0+H^2{\cal S}'
          \left[\frac{2-5c}{1-c}(x\cdot\theta')\right.\\
           \\\;\;\;\;\;\;\;\;\;\;\;\;\;\;\;\;\;\;\;\;\;\;\;\;\left.+\frac{c}{1-c}\left(D_1\theta'\cdot 
+x\theta'\cdot-H^{-2}\theta'\cdot\bar{\partial}\right)\right]{\cal
W}_{1},\;\;\;$(III)$
        \end{array}\right.\e
where the condition $\partial.{\cal W}_{1}=0$ is used. By imposing the following condition \b \theta'{\cal
W}_0(x,x')=-\frac{2}{3}{\cal S}'\theta'\cdot{\cal W}_{1}(x,x') ,\e  the bi-tensor two-point function $(4.54)$ can be written explicitly in terms of bi-vector two-point function ${\cal W}_1$. The
bi-vector two-point function ${\cal W}_{1}$ can be written in the
following form \cite{gata00,gagarota08}
$${\cal W}_{1}=\theta\cdot\theta'{\cal W}_{2}+D_1D'_1{\cal W}_{3},$$
where ${\cal W}_{2}$ and ${\cal W}_{3}$ are bi-scalar two-point
functions. They satisfy the following equations
$$D'_1{\cal W}_{3}=-\frac{1}{2}\left[2H^2(x\cdot\theta'){\cal W}_{2}-\theta'\cdot\bar{\partial}{\cal W}_{2}\right],$$
$$Q_0{\cal W}_{2}=0.$$
This means, ${\cal W}_{2}$ is a massless minimally coupled
bi-scalar two-point function. Putting ${\cal W}_{2}={\cal
W}_{m},$ we have \b {\cal W}_{1}(x,x')=\left(\theta\cdot\theta'
    -\frac{1}{2}D_{1}[\theta'\cdot\bar \partial+2H^2
x\cdot\theta']\right){\cal W}_{m}(x,x').\e

Using $(4.56)$ in $(4.55$-III), we have
$$(Q_1+6)D'_2{\cal W}_g=\frac{cH^2}{1-c}H^2{\cal S}'
\left[\frac{2-5c}{c}(x\cdot\theta'){\cal
W}_{1}+\frac{1}{3}D_1(\theta'\cdot{\cal W}_{1})+x(\theta'\cdot{\cal
W}_{1})-H^{-2}\theta'\cdot\bar{\partial}{\cal W }_{1}\right].$$ By using
the following identities \cite{gata00,derotata08}
$$(Q_0+6)^{-1}(x\cdot\theta'){\cal W}_{1}=\frac{1}{6}
\left[\frac{1}{9}D_1(\theta'\cdot{\cal W}_{1})+(x\cdot\theta'){\cal
W}_{1}\right],$$
$$(Q_1+6)\theta'\cdot\bar\partial
{\cal W}_{1}=6\theta'\cdot\bar\partial {\cal
W}_{1}+2H^2D_1(\theta'\cdot{\cal W}_{1}),$$
$$(Q_1+6)D_1\theta'\cdot{\cal W}_{1}=6D_1(\theta'\cdot{\cal W}_{1}),$$
$$(Q_1+6)x\theta'\cdot{\cal W}_{1}=6x (\theta'\cdot{\cal W}_{1}),$$  one can obtain
\b D'_2{\cal W}_g(x,x')=\frac{cH^2}{6(1-c)} {\cal S}'
\left[\frac{2+c}{9c}D_1\theta'\cdot{\cal W}_{1}
+\frac{2-5c}{c}x\cdot\theta'{\cal W}_{1}+x\theta'\cdot{\cal W}_{1}
-H^{-2}\theta'\cdot\bar \partial{\cal W}_1\right].\e

By using equations $(4.56-58)$, it turns out that the bi-tensor
two-point function can be written in the following form  $(c=\frac{2}{5})$: \b {\cal
W}_{\alpha\beta \alpha'\beta'}(x,x')=\Delta_{\alpha\beta
\alpha'\beta'} (x,x'){\cal W}_{m}(x,x'), \e where
$$ \Delta(x,x')=-\frac{2}{3}{\cal S'}\theta
\theta'\cdot\left(\theta\cdot\theta'
    -{\frac{1}{2}}D_{1}[2H^2 x\cdot\theta'+\theta'\cdot\bar\partial]\right)$$
$$ +{\cal S}{\cal S}'\theta\cdot\theta'\left(\theta\cdot\theta'
    -{\frac{1}{2}}D_{1}[2H^2 x\cdot\theta'+\theta'\cdot\bar\partial]\right)$$ \b +\frac{H^2}{9} {\cal
S}'D_2 \left(\frac{2}{3}D_1\theta'\cdot + x\theta'\cdot -
H^{-2}{\theta'}\cdot{\bar\partial}\right)\left(\theta\cdot\theta'
    -{\frac{1}{2}}D_{1}[2H^2 x\cdot\theta'+\theta'\cdot\bar\partial]\right),\e
and ${\cal W}_{m}$ is the two-point function for the
minimally coupled scalar field. The bi-scalar two-point function ${\cal W}_{m}$ in the ``Gupta-Bleuler vacuum''
state is \cite{ta3} \b {\cal W}_{m}(x,x')=\frac{iH^2}{8\pi^2}
\epsilon (x^0-x'^0)[\delta(1-{\cal Z}(x,x'))+\vartheta ({\cal
Z}(x,x')-1)],
\e where $\vartheta$ is the Heaviside step function and \b \epsilon (x^0-x'^0)=\left\{ \ba{rcl} 1&x^0>x'^0 ,\\
0&x^0=x'^0 ,\\ -1&x^0<x'^0.\\ \ea\right.\e 
 $ {\cal{Z}}$  is an invariant object under the
isometry group $O(1,4)$. It is defined for $x$ and $x'$ on the
dS hyperboloid by: $$ {\cal{Z}}\equiv
-x\cdot x'=1+{\frac{1}{2}}(x-x')^2.$$

\subsection{Tensor Field operator}

Detailed knowledge of the two point function, ${\cal W}_{\alpha\beta \alpha'\beta'}(x,x')$, enable us to construct quantum
field operators. The tensor fields $\K(x)$ is expected to be
an operator-valued distribution on $X_H$. Note that $\K(x)$ acts on a complex
vector space ${\cal V}$ with an indefinite metric. In terms of
complex vector space and field-operator, the properties of the two-point function ${\cal W}$ are equivalent to the following properties:
\begin{enumerate}

\item {\bf Existence of an indecomposable representation of
the dS group}.

\item {\bf Existence of at least one ``vacuum state''} $|GB>$, cyclic for the polynomial algebra of field operators and invariant
under the indecomposable representation of dS group.

\item {\bf Existence of a complex
vector space} ${\cal V}$ with an indefinite sesquilinear form that
 can be described as the direct sum
\b {\cal V}={\cal V}_0 \bigoplus[\bigoplus_{n=1}^{\infty}S{\cal
V}_1^{\bigotimes n}].\e $S$ denotes the symmetrization operator
and ${\cal V}_0=\{ \lambda |GB>,\;\; \lambda \in \C\}$.
 ${\cal V}_1\equiv V_c$ is defined with the indefinite sesquilinear form
\b (\Psi_1,\Psi_2)=\int_{X_H \times X_H}
\Psi_1^{*\alpha\beta}(x){\cal W}_{\alpha\beta \alpha'\beta'}
(x,x')\Psi_2^{\alpha'\beta'}(x')d\sigma(x)d\sigma(x'),\e where
$\Psi_{\alpha \beta} \in {\cal D}(X_H)$.
\item {\bf Covariance}, the field operator can be transformed by an indecomposable representation of dS group,
 \b U(g) \K_{\alpha\beta}(x) U(g^{-1})=
g_{\alpha}^{\gamma} g_{\beta}^{\delta}\K_{\gamma \delta}(g x).\e

\item{\bf Locality} for every space-like separated pair $(x,x')$ \b
[\K_{\alpha\beta}(x),\K_{\alpha'\beta'}(x')]=0. \e
\item {\bf Transversality}
\begin{equation} x\cdot \K(x)=0.\end{equation}
\item {\bf Index symmetrizer}
\begin{equation}  \K_{\alpha \beta}=\K_{\beta
\alpha}.  \end{equation}
\item {\bf Tracelessness}
\begin{equation}  \K_{\alpha }^{\;\;\alpha}=0.  \end{equation}
\end{enumerate}

We now define the field operator that satisfies the above properties and provides the two-point
function $(4.59)$. Using the Eqs.
$(4.41)$ and $(2.8)$, the field operator in Krein space is defined as $$
\K_{\alpha \beta}(x)=\sum_{\lambda Llm}
a_{Llm}^{\lambda}{\cal
E}_{\alpha\beta}^{\lambda}(\rho,\Omega,Llm)\phi_{Llm}(\rho,\Omega)+H.C.$$ \b
+\sum_{\lambda Llm} b_{Llm}^{\lambda}\left[{\cal
E}_{\alpha\beta}^{\lambda}(\rho,\Omega,Llm)\phi_{Llm}(\rho,\Omega)\right]^*+H.C.,\;\;\;\;\forall \;0\leq l \leq L, \;\;\; -l\leq m
\leq l. \e The Gupta-Bleuler vacuum is defined by \b
a_{Llm}^{\lambda}|GB>=0,\;\; b_{Llm}^{\lambda}|GB>=0.\e
The commutation relation between the annihilation and creation operators are:
$$[a_{Llm}^{\lambda},a_{L'l'm'}^{\dag \lambda '}]=f(\lambda)\delta_{\lambda
\lambda '}\delta_{LL'}\delta_{ll'}\delta_{mm'},$$
$$[b_{Llm}^{\lambda},b_{L'l'm'}^{\dag \lambda '}]=-f(\lambda)\delta_{\lambda
\lambda '}\delta_{LL'}\delta_{ll'}\delta_{mm'},$$ where
$f(\lambda)$ is a sign function (positive or negative) defined as: \b f(\lambda)\equiv\left\{ \ba{rcl}1 \;\;\;\;\; \mbox{for}\;\;\;\;\;\; \lambda=1,...,6 ,\\ -1 \;\;\;\;\;\;  \mbox{for}\;\;\; 
\lambda=7,...,10.\\ \ea\right.\e 

In this case, we have $10$ polarization states for $a$ modes and $10$ polarization states for $b$ modes, amongst which two of them
are physical (transverse traceless positive frequency mode). These modes can be defined by: $$ (a_{Llm}^{\lambda})^\dagger|0>=|1_{Llm}^{\lambda (a)}>\equiv|\mbox{physical state}>,\;\; \lambda=1,2.$$
There are two type of un-physical states. The first type is the usual mode, which appear due to the gauge invariance of the field equation. These un-physical positive frequency modes are: $$ (a_{Llm}^{\lambda})^\dagger|0>=|1_{Llm}^{\lambda (a)}>\equiv|\mbox{unphysical state}>,\;\; \lambda=3,....,8,$$ four of which have negative norms. The other un-physical states are due to the Krein space quantization with negative frequency mode:
$$(b_{Llm}^{\lambda})^\dagger|0>=|1_{Llm}^{\lambda (b)}>\equiv|\mbox{unphysical negative frequency state}>,
\;\; \lambda=1,..,10.$$  Let us insist here that the Krein
procedure allows us to avoid the pathological large-distance
behaviour of the graviton propagator and preserves the de Sitter invariant.

\section{Conclusion and outlook}

In this construction, even though we have introduced the
artificial device of Krein space quantization, the
physical quantities always refer to states with positive frequency. We conclude that with the conditions i) and ii) in page $3$, the un-physical states do not contribute to the S matrix elements, so unitarity preserve. Even thought un-physical states have disappeared from the physical subspace, their impact on automatic regularization of two-point function remains as an excellent tool for resolving the problem of infra-red divergences. 

This construction although successfully removes the infra-red divergence and preserves the de Sitter covariance for quantum linear gravity, does not maintain the analyticity of the two-point function. Another problematic issue in this approach is the difference between the introduced vacuum state (Gupta-Bleuler vacuum) with the Bunch-Davies vacuum state which commonly use for QFT in de Sitter space. In the forthcoming paper ({Quantum Linear Gravity in de Sitter Universe On Bunch-Davies vacuum state), these obstacles are successfully removed \cite{taro14}.   

\vspace{0.5cm} \noindent {\bf{Acknowlegements}}: We are grateful
to J. Iliopoulos and J.P. Gazeau for their helpful discussions and S. Teymourpoor for her interest in this work.

\end{document}